\documentclass[
reprint,
superscriptaddress,
amsmath,amssymb,
aps,
floatfix
]{revtex4-2}
\usepackage{ragged2e}
\usepackage{graphicx}
\usepackage{dcolumn}
\usepackage{bm}
\usepackage{float}

\usepackage{xcolor}

\begin{document}

\setcitestyle{super}
\preprint{APS/123-QED}
\title{
Machine learning the relationship between Debye temperature and superconducting transition temperature
}

\author{Adam D. Smith}
\email{smitha20@uab.edu}
\affiliation{Department of Physics, University of Alabama at Birmingham, Birmingham, Alabama 35294, USA}

\author{Sumner B. Harris}
\affiliation{Center for Nanophase Materials Sciences, Oak Ridge National Laboratory, Oak Ridge, Tennessee 37831, USA}

\author{Renato P. Camata}
\affiliation{Department of Physics, University of Alabama at Birmingham, Birmingham, Alabama 35294, USA}

\author{Da Yan}
\affiliation{Department of Computer Science, University of Alabama at Birmingham, Birmingham, Alabama 35294, USA}

\author{Cheng-Chien Chen}
\email{chencc@uab.edu}
\affiliation{Department of Physics, University of Alabama at Birmingham, Birmingham, Alabama 35294, USA}

\date{\today}
\begin{abstract}
Recently a relationship between the Debye temperature $\Theta_D$ and the superconducting transition temperature $T_c$ of conventional superconductors has been proposed [npj Quantum Materials {\bf 3}, 59 (2018)]. The relationship indicates that $T_c \le A \Theta_D$ for phonon-mediated BCS superconductors, with $A$ being a pre-factor of order $\sim 0.1$. In order to verify this bound, we train machine learning (ML) models with 10,330 samples in the Materials Project database to predict $\Theta_D$. By applying our ML models to 9,860 known superconductors in the NIMS SuperCon database, we find that the conventional superconductors in the database indeed follow the proposed bound.
We also perform first-principles phonon calculations for H$_{3}$S and LaH$_{10}$ at 200 GPa. The calculation results indicate that these high-pressure hydrides essentially saturate the bound of $T_c$ versus $\Theta_D$.
\end{abstract}

\maketitle
\section{Introduction}
The Eliashberg theory~\cite{eliashberg1960interactions, eliashberg1961temperature, marsiglio2020eliashberg} describes frequency-dependent phonon-mediated attractive interaction between electron Cooper pairs, by taking into account phonon dynamics and retardation effects. A key quantity in the theory is the Eliashberg spectral function $\alpha^2 F(\omega)$, which is related to the electronic density of states (DOS) near the Fermi level, the phonon spectra, and the electron-phonon (el-ph) coupling matrix elements. 
The Eliashberg function enables the determination of the total (isotropic) el-ph coupling parameter $\lambda$, which together with the effective Coulomb pseudopotential $\mu^*$ (typically chosen to be between $\sim 0.1-0.15$) can be utilized to estimate the $T_c$ of conventional superconductors, via analytical expressions such as the McMillan~\cite{mcmillan1968transition}, Allen-Dynes~\cite{allen1975transition}, or other related formula~\cite{xie2022machine}, obtained by fittings to numerical solutions of the Eliashberg equations.

The Bardeen-Cooper-Schrieffer (BCS) theory~\cite{bardeen1957microscopic, bardeen1957theory} dictates that $T_c = 1.13\, \omega_c e^{-\frac{1}{\lambda}}$~\cite{}, with $\omega_c$ a characteristic phonon frequency. 
In the weak coupling limit of small $\lambda$, the Eliashberg theory $T_c$ essentially reduces to the BCS $T_c$ (up to a modified pre-factor~\cite{carbotte1990properties,marsiglio2018eliashberg}). In the strong coupling limit of large $\lambda$, it was shown that $T_c \propto \sqrt{\lambda}$ in the Eliashberg theory~\cite{allen1975transition}.
On the other hand, when $\lambda$ becomes very large, the characteristic phonon frequency can be substantially re-normalized (softened)~\cite{allen1972superconductivity, gomersall1974variation, zhang2005nonlocal, moussa2006two, schrodi2021influence}, leading to a decrease in $T_c$.
Moreover, in the presence of a significantly large $\lambda$, superconductivity can be suppressed by other instabilities such as charge density wave or by the formation of heavy polarons (localized charge carrier dressed by phonons)~\cite{macridin2006synergistic, ohgoe2017competition, karakuzu2017superconductivity, bradley2021superconductivity, feng2022phase}, in which cases the $T_c$ predicted by the Eliashberg theory is not applicable. 
More recently, determinant quantum Monte Carlo (DQMC) studies of two-dimensional Holstein models~\cite{esterlis2018breakdown, esterlis2019pseudogap, chubukov2020eliashberg, nosarzewski2021superconductivity} have shown that the Eliashberg theory might already fail when $\lambda$ approaches a critical value $\lambda_{cr} \sim O(1)$, {\it{without}} the formation of any instability. 
In these regards, $T_c$ would remain bounded even if $\lambda$ can assume an arbitrarily large value.

Based on the DQMC results and the relationship between $T_c$ and the Debye temperature $\Theta_D$ for a few tens of selected compounds, Esterlis {\it et al.} have proposed that $T_c \le A \Theta_D$ for phonon-mediated superconductors~\cite{esterlis2018bound}.
Here, $A \sim 0.1$ is a material-independent upper limit, while the actual material-specific ratio $T_c/\Theta_D$ is determined by details of the band structure, el-ph coupling strength, Coulomb interaction, etc.
If the above relationship is generically applicable to all BCS superconductors, it has several important implications: (i) For a material whose $T_c$ is much smaller than the bound, engineering el-ph or other properties might further enhance $T_c$; (ii) Near-room-temperature BCS superconductors (with $T_c \gtrsim 250 $ K) may be achieved in materials with $\Theta_D \gtrsim 2,500$ K, which can typically occur in light-element compounds under high pressure~\cite{peng2017hydrogen, drozdov2019superconductivity, somayazulu2019evidence}.
Therefore, it is an important task to examine the proposed relationship between $T_c$ and $\Theta_D$ for more known BCS superconductors.

In this study, we employ data-driven approaches to achieve the above task.
Specifically, we develop machine learning (ML) models to predict $\Theta_D$, and then compare the resulting $\Theta_D$ to the experimental $T_c$ in the SuperCon database~\cite{Supercon:2022}.
Manual selection criteria and machine learning classification techniques are further utilized to separate conventional superconductors from others in the database. We show that for nearly 2,000 conventional superconductors, they all satisfy the bound of Esterlis {\it et al.}~\cite{esterlis2018bound} 
Moreover, we perform first-principles phonon calculations to investigate H$_3$S and LaH$_{10}$ under 200 GPa, and find that $T_c \sim 0.1$ to 0.11 $\Theta_D$. This result indicates that these high-pressure hydrides have essentially saturated the proposed bound. 
Our study thereby suggests that while it might be possible to slightly fine tune the pre-factor $A$ by engineering material properties via external perturbations, it appears {\it unlikely} that $T_c$ can increase to a value much larger than $0.1\Theta_D$ in BCS superconductors.


\section{Computational Methods}
\subsection{Machine Learning Models}

\textit{Data Acquisition.}
The open Materials Project (MP) database~\cite{Jain2013} contains calculated properties of over 145,000 materials. At the time of sourcing, 12,370 crystals were present with documented elastic tensors in MP.
We use the Pymatgen package~\cite{Ong2013,Ong2015} to interface with MP and downselect 10,330 compounds to ensure that the training dataset includes only samples with adequate mechanical properties.
In particular, we first remove compounds with elastic tensors which have negative eigenvalues (indicative of mechanical instability), or with elastic constants $C_{12}$ and $C_{13}$ greater than $C_{11}$ (which results in non-physical calculations for the bulk and shear moduli).
We also remove structures that potentially exist only under high pressure, which is achieved by excluding compounds that have a unit-cell volume per atom less than that of cubic diamond (which has the known smallest unit-cell volume per atom at ambient pressure).
Finally, we remove 340 compounds that also exist in the superconductor dataset discussed below. This set of 340 materials are used later for an unbiased evaluation of the machine learning (ML) models (see the Results section).

The superconductor dataset under study is sourced from the Japanese National Institute of Material Science (NIMS) SuperCon database~\cite{Supercon:2022}. This database features over 30,000 superconducting compound entries with elemental composition, experimental $T_{c}$, and journal references. Many of these compounds have additional information such as their structural likeness (ABO3, Y123, NaCl ...), lattice constants, and multiple measurements with unique journal references. 
We first downselect from the entire SuperCon database to samples with known structural-likeness (from which we derive the crystal system information as an ML model input feature).
Moreover, compounds with multiple measurements of $T_c$ (from different journal references) that deviate by $<$10\% are consolidated into one sample with a simple-averaged $T_c$, while those with $>$10\% deviation are removed entirely.
The above downselection procedure results in a dataset containing 9,860 unique superconducting compounds with $T_{c}$ and crystal symmetry group information. This dataset with our ML predicted density and Debye temperature information is stored in a JSON file downloadable at \url{github.com/condmatr/Debye-ML/tree/main/superconductor_dataset}.

\textit{Feature Generation.}
The ML training features are generated by using the MatMiner package~\cite{Ward2018}. We first generate features derivable from the composition, including elemental fraction (for atoms H to Lr), elemental statistics (mean and range for atomic mass, periodic table column, periodic table row, atomic number, atomic radii, and electronegativity), statistics for valence electrons (mean valence electrons, and fraction of total valence electrons in $s$, $p$, $d$, and $f$ orbitals), according to Meredig \textit{et al.}~\cite{meredig2014combinatorial} Additionally, we include the transition metal fraction (also derived from the composition), and one-hot encoded crystal system (derived from the crystal structure), as these are all congruent features existing in the SuperCon dataset.
One hot encoding converts categorical information into a numerical format for training machine learning models. For example, each category value is converted into a new feature vector with element values being 1 or 0, respectively for the presence or absence of a specific crystal system.
In total, we use 128 features to predict the density, and 129 features (including additionally the predicted density) to predict the Debye temperature. These features and the target data for training the ML models are saved in a JSON file downloadable at \url{github.com/condmatr/Debye-ML/tree/main/model_training_dataset}.

\textit{Model Training, Validation, and Application.}
Our regression models for the density and Debye temperature are based on gradient-boosted trees, which is an ensemble algorithm that builds decision trees sequentially and aims at reducing the error of the previous tree at each iteration.
In particular, we utilize the XGBoost package~\cite{chen2016xgboost}, which has one of the best implementations of gradient boosted trees algorithm due to its great performance for regression and classification problems.
We also test other ensemble tree methods such as the random forests implemented in Scikit-learn~\cite{scikit-learn, sklearn_api}, and find that the models based on XGBoost show superior performance (at the cost of not being parallelizable, since the tree is gradient boosted one round at time), which is a result of additive tree-building and post-tree pruning.

To train the ML models, we use grid search  (with over 4000 permutations) to tune the hyperparameters, including the maximum tree depth, minimum samples per node, ridge and lasso regularization coefficients, and loss-based post-tree pruning. 
The tuning process is conducted with a 10-fold cross-validation on a 85\% training set randomly split from the total dataset (with the remaining 15\% being the test set).
To prevent overfitting, we also limit the minimum amount of samples per tree node to 30 and the maximum tree depth to 6, which leads to models that generalize better to new data. The Python code needed to create the model features and implement our models is available at \url{github.com/condmatr/Debye-ML}.

As discussed in the Results section, we impose manual selection criteria based on the elemental composition to separate BCS superconductors from the unconventional ones. To check that our criteria are sufficient to separate different classes of superconductors, we further develop a classification model based on Scikit-learn's implementation of linear discriminant analysis (LDA)~\cite{scikit-learn, sklearn_api}. 
LDA is a supervised ML classification model, and it accomplishes the training by calculating intra-class and inter-class scatter matrices for the set of samples with pre-assigned class labels, and uses features of those samples to make new (LD) axes created by linear combinations of the features to separate the classes in the latent space. 
We apply LDA to samples in the superconductor dataset belonging to different class labels, including the ``Cu and O" (cuprates), ``iron-based" superconductors, ``lanthanide and actinide" superconductors, and ``Others". We then score the classification by computing a class-weighted $F_{1}$ score (see the Results section).

\subsection{First-Principles Calculations}
First-principles calculations are based on density functional theory (DFT) using the Vienna Ab initio Simulation Package (VASP)~\cite{Kresse1996,Kresse19962}. The Monkhorst–Pack sampling scheme~\cite{Monkhorst1976} is used with a $\Gamma$-centered k-point  mesh, having a sampling-point resolution of $0.02 \times 2 \pi$ per \AA. The convergence criteria of self-consistent and structural relaxation calculations are set to 10$^{-6}$ eV per unit cell and 10$^{-3}$ eV per \AA, respectively. We adopt a planewave cutoff energy of 363 eV, which is sufficient to converge the DFT total energy difference within $10^{-4}$ eV per atom for both H$_{3}$S and LaH$_{10}$. All calculations use the projector augmented wave (PAW)~\cite{Blochl1994,Kresse1999} pseudopotentials and the Perdew–Burke–Ernzerhof generalized gradient approximation (GGA-PBE) functional\cite{Perdew1996}. Phonon spectra are computed using the PHONOPY package\cite{Togo2015} with the finite displacement method on $4 \times 4 \times 4$ supercells. Adequate convergences of the force constants and the phonon dispersion for both systems are carefully checked.

\section{Results \& Discussion}
\subsection{Machine Learning of Debye Temperature}

There are various ways to determine the Debye temperature $\Theta_D$, which is in general related to a material's phonon and mechanical properties.
For the purpose of training machine learning (ML) models with sufficient target samples, we choose to compute $\Theta_D$ from the mean sound velocity $v_m$ using the method by Anderson \textit{et al.}~\cite{ANDERSON1963909}:

\begin{equation}\label{eq:1}
\Theta_D = \frac{\hbar}{k_B}\left[ \frac{6\pi^2 q}{V_0}\right]^\frac{1}{3} v_m,
\end{equation}
where $q$ is the number of atoms in the unit cell, and $V_0$ is the unit-cell volume. $v_{m}$ can be further expressed in terms of the longitudinal and transverse sound velocities, $v_l$ and $v_t$, respectively:
\begin{equation}\label{eq:2}
v_m = \left( \frac{1}{3} \left(\frac{2}{v^3_t} + \frac{1}{v^3_l} \right)\right)^{-\frac{1}{3}}.
\end{equation}
$v_l$ and $v_t$ in turn are related to the mechanical properties:
\begin{equation}\label{eq:3}
v_t = \sqrt{\frac{G}{\rho}}, \,\,\, v_l = \sqrt{\frac{B + \frac{4}{3}G}{\rho}}.
\end{equation}
Here, $G$ and $B$ are respectively the shear and bulk moduli; $\rho$ is the density of the crystal.
We note that $\Theta_D$ computed by the above method is consistent with those determined from phonon calculations and experiments (see TABLE I for a comparison).

\begin{table}[t!]
\centering
\begin{tabular}{c c c c} 
\hline\hline
\quad Compound \quad & \quad $\Theta_{D,v}$ \quad  & \quad $\Theta_{D,p}$ \quad & \quad $\Theta_{D,e}$ \quad \\ [0.5 ex]
\hline
MgB$_{2}$ & \quad 1025K \quad \cite{Wang2015} & \quad 1044K \quad \cite{Bohnen2001} & \quad 920K \quad \cite{WANG2001}\\ [0.5 ex]
\quad C (cubic diamond) \quad & \quad 2245K \quad \cite{Field_2012} & \quad 2240K \quad\cite{tohei2006} & \quad 2240K \quad \cite{gschneidner1964physical}\\[0.5 ex]
\hline\hline
\end{tabular}
\caption{Debye temperatures of MgB$_{2}$ and C (cubic diamond) determined from the sound velocity ($\Theta_{D,v}$), phonon density of states ($\Theta_{D,p}$), and experiment ($\Theta_{D,e}$), respectively.}
\label{table:features}
\end{table}

\begin{figure}[h!]
    \includegraphics[width=0.47\textwidth]{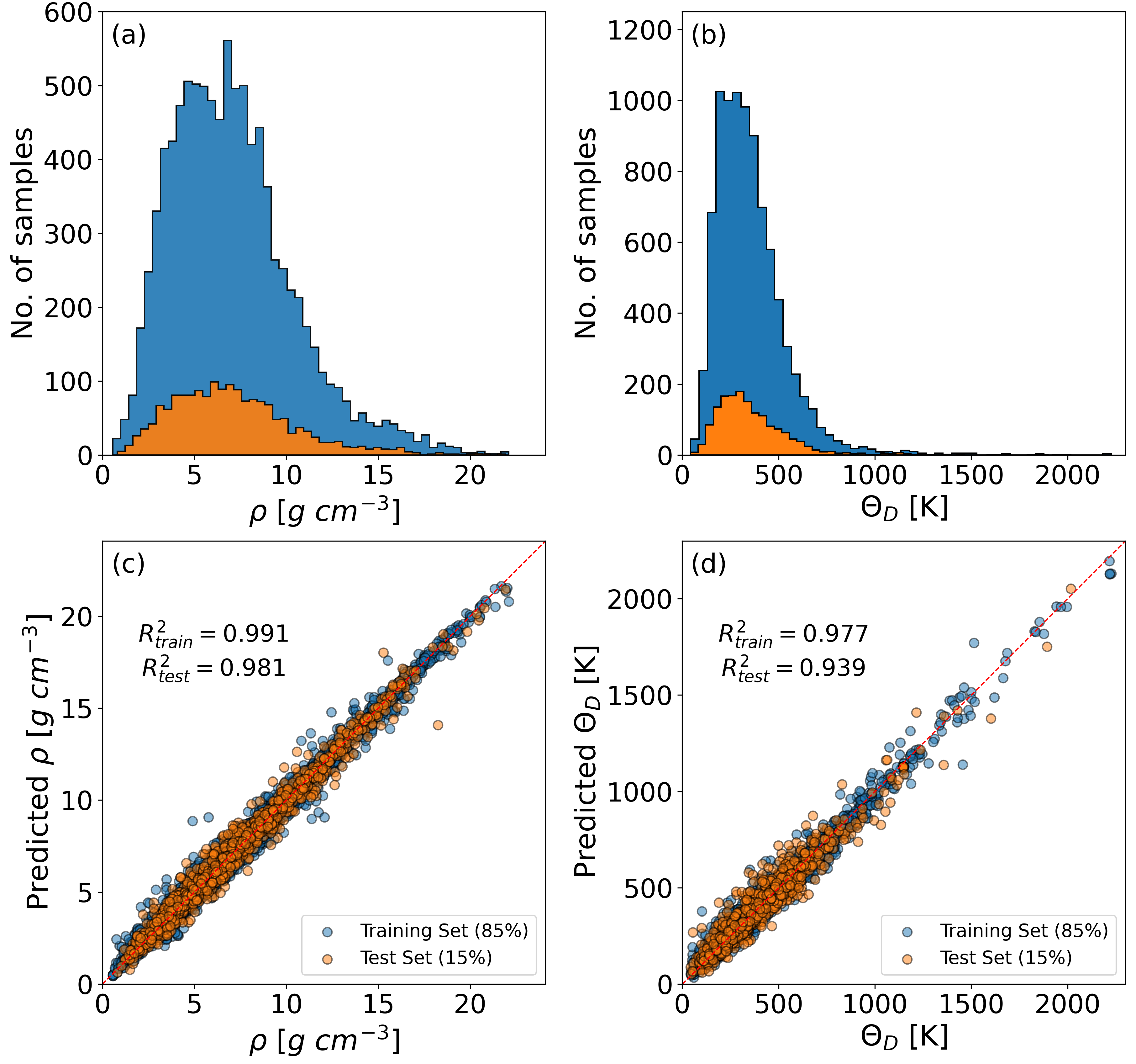}
    \caption{
    Training data distribution and model validation for gradient-boosted trees on 10,330 compounds from the Materials Project database.
    \textbf{(a), (b)} Histograms of the training sets (blue color) and test sets (orange color) respectively for the targets of density $\rho$ and Debye temperature $\Theta_D$.
    \textbf{(c), (d)} Scatter plots between the machine learning predictions (on the $y$-axes) and the actual values (on the $x$-axes) respectively for $\rho$ and $\Theta_D$. The coefficient of determination $R^2$ score is used as the evaluation metric. Both models achieve high $R^2$ scores in the training and test sets.
    }
    \label{fig:model_training}
\end{figure}

Compared to lattice dynamics information such as phonon spectra, static mechanical properties are more commonly available in open materials database, which thereby facilitates training ML models. In particular, we consider 10,330 training samples downselected from the Materials Project (MP)~\cite{Jain2013} for ambient pressure condition (see the Methods section), obtain their density information $\rho$, and extract the $G$ and $K$ values based on the elastic modulus tensors in MP. The final target data of $\Theta_D$ for each training sample is then determined by Eq.~\ref{eq:3}.
Figures \ref{fig:model_training}(a) and \ref{fig:model_training}(b) show respectively the resulting histogram distributions of $\rho$ and $\Theta_D$ for our 10,330 training samples, randomly split into a 85\% training set (blue color) and a 15\% test set (orange color).

With the target data, we train two separate ML models using gradient-boosted trees to predict respectively $\rho$ and $\Theta_D$. For the density, we generate 128 features based on elemental fraction, compositional statistics, electronic statistics, and one-hot encoded crystal system, as discussed in the Methods section. For the Debye temperature, we adopt the density as an additional feature (resulting in a total of 129 features) to improve further the model performance for predicting $\Theta_D$.
To tune the hyperparameters of the gradient boosted trees, we use a combined grid searches and cross validations on the training data set (see the Methods section for further details).

Figures \ref{fig:model_training}(c) and \ref{fig:model_training}(d) show the resulting model performances for predicting $\rho$ and $\Theta_D$, respectively.
For model evaluation, we use the associated coefficient of determination $R^2$ score, which ranges between 0 (meaning no linear relationship between the predicted and actual values) to 1 (meaning 100\% accuracy in the prediction).
The final models achieve high scores of $R^{2}=0.981$ and $R^{2}=0.939$ on the test sets for predicting $\rho$ and $\Theta_D$, respectively.
Among the 129 features used in the Debye temperature model, the most important (non-elemental) features are mean atomic weight and mean atomic number, which are both negatively correlated with $\Theta_D$; that is, compounds with low mean atomic mass and low mean atomic number tend to exhibit a high Debye temperature, as anticipated.

\begin{figure}[h!]
    \centering
    \includegraphics[width=0.48\textwidth]{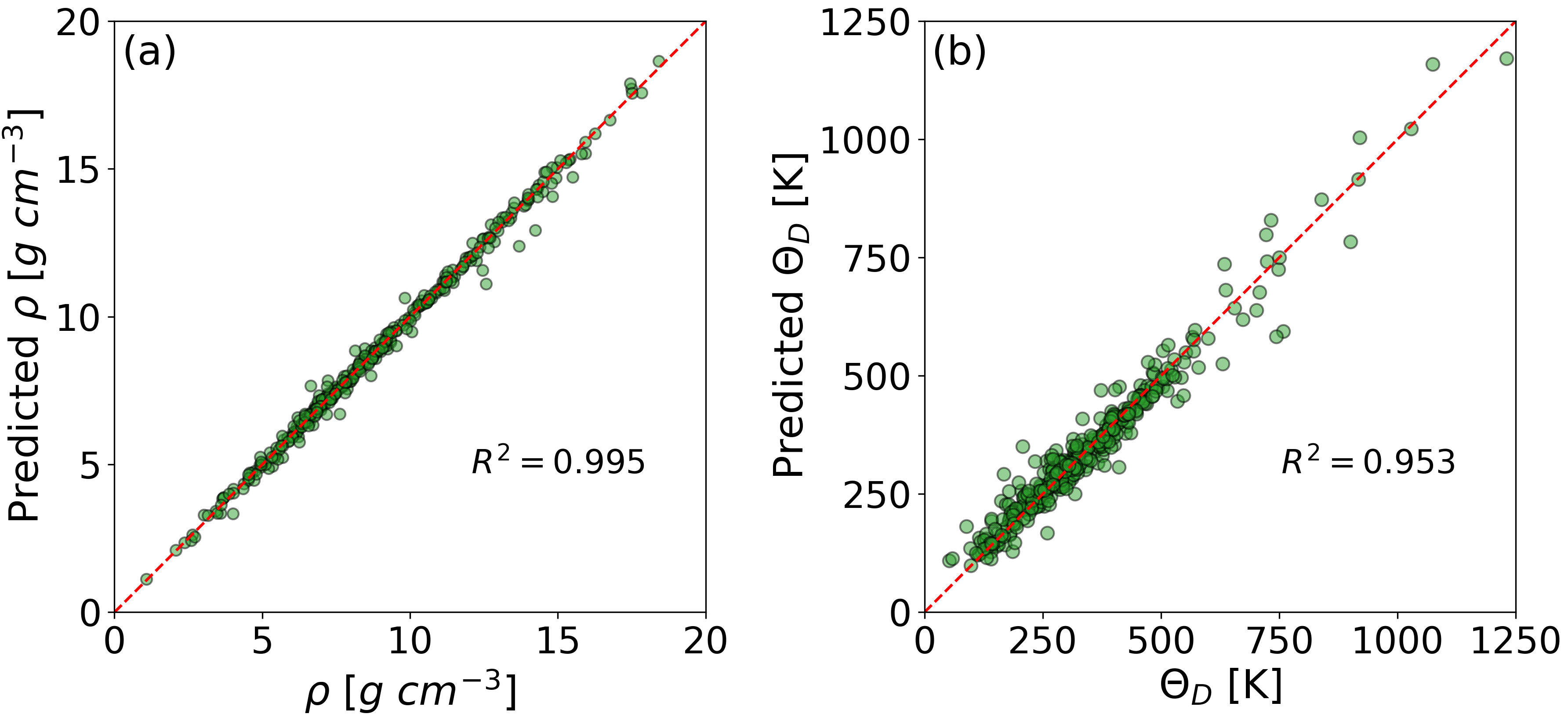}
    \caption{
    Evaluation of the density $\rho$ and Debye temperature $\Theta_D$ machine learning models on 340 materials existing in both the Materials Project and the SuperCon databases. These materials were never seen by our models during the training. The models achieve highly accurate predictions, with respective $R^2$ scores (between the predicted and actual values) of 0.995 for $\rho$ and 0.953 for $\Theta_D$.
    }
    \label{fig:Supercon_model_validation}
\end{figure}

To further test our ML models, we apply them to 340 materials that exist in both the Materials Project and the SuperCon databases. We note that these 340 superconducting compounds were first removed on purpose from our target datasets prior to training the regression models for an unbiased evaluation. 
As shown in Fig. \ref{fig:Supercon_model_validation}, our regression models achieve $R^2$ scores of 0.995 and 0.953 respectively for predicting $\rho$ and $\Theta_D$. These results indicate that the ML predictions are \textit{highly accurate even on samples never seen by our models before}. 

\subsection{Classification of Superconductors}

The superconductors under study are sourced from the SuperCon database \cite{Supercon:2022}, from which we downselect 9,860 unique superconducting compounds with $T_c$ and symmetry-group information (see the Methods section). The selected compounds in general consist of different superconductor families, and we proceed by separating the compounds into different classes, such as the cuprates, iron pnictides and chalcogenides, heavy-fermion superconductors, and conventional phonon-mediated superconductors.
We first use a manual selection based on the elemental composition for classification. In particular, we iteratively sort out compounds containing elements Cu and O (the cuprates), then compounds containing Fe but not Cu (iron-based superconductors), and finally compounds containing lanthanide or actinide elements (heavy-fermion superconductors); the remaining are labeled as ``Others", which in principle contain mostly conventional BCS superconductors.
This manual classification is admittedly not 100\% accurate. For example, some unconventional superconductors like SrTiO$_3$ may be classified into the ``Others" category. Additionally, there are cases where lanthanide and actinide compounds are more likely BCS superconductors, such as LaH$_{10}$ (present in our data), which are typically only stable under high pressure. We note that our ML model trained mainly on ambient-pressure data does not take pressure as an input feature. Therefore, the ML predicted $\Theta_D$ for the hydrides would be erroneously low. We will address the high-pressure hydrides directly via first-principles calculations discussed later in this section.

\begin{figure}[h!]
    \includegraphics[width=0.47\textwidth]{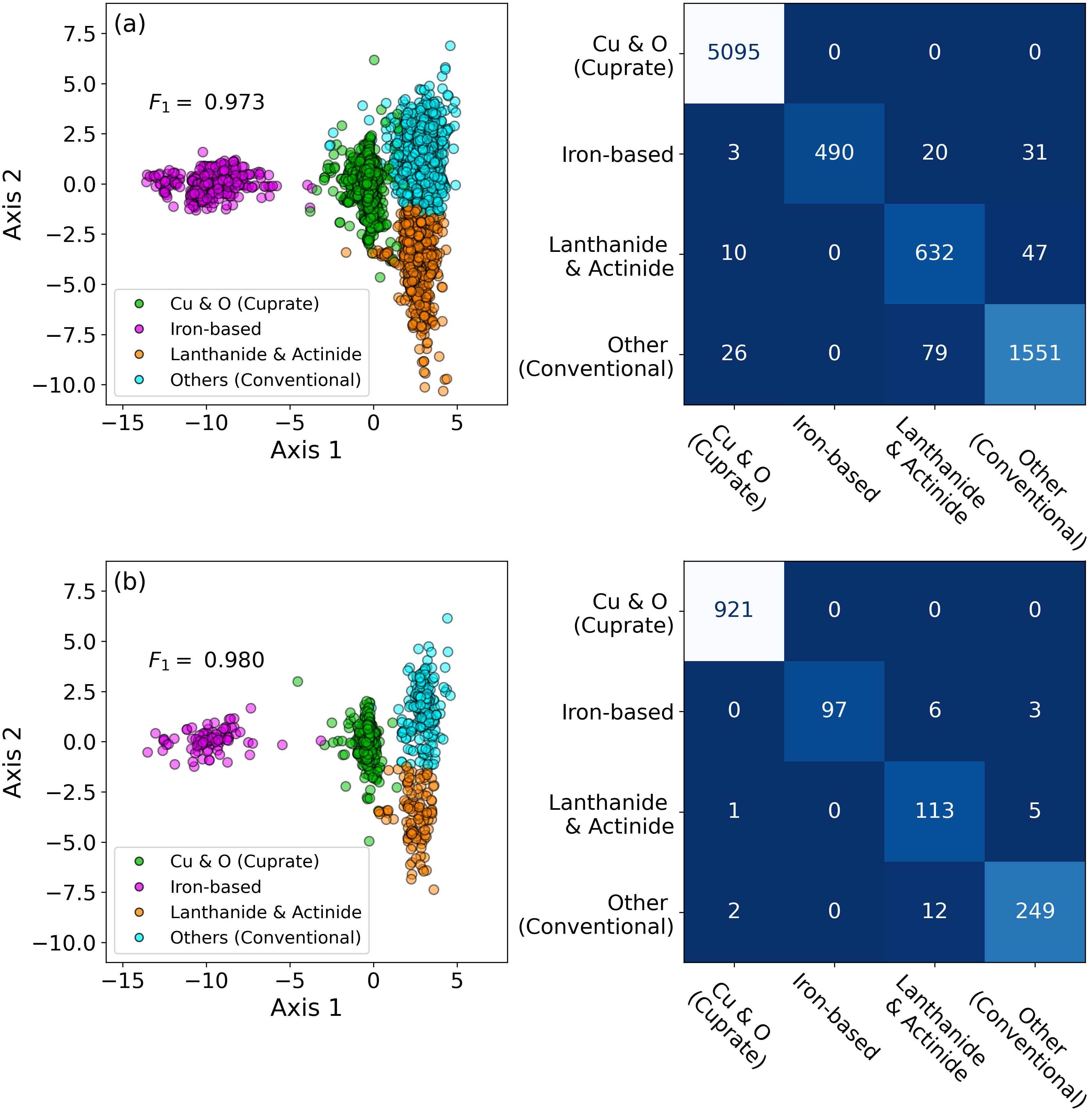}
    \caption{
    Linear discriminant analysis (LDA) classification of superconductors for the \textbf{(a)} training set (85\%) and \textbf{(b)} test set (15\%). The LDA model achieves high class-weighted F$_1$ scores of 0.973 and 0.980 respectively for the training and test sets.
    The left columns show that different superconductor families are well separated in the latent space spanned by the first and second LD axes. The right columns show that the confusion matrices (with the $x$-axis the predicted label and the $y$-axis the true label) have predominant diagonal elements, indicating a majority of correction classification predictions.
    }
    \label{fig:LDA}
\end{figure}

To further test our manual classification scheme, we employ the linear discriminant analysis (LDA) technique~\cite{Li2006}. LDA is generally capable of constructing an $\mathbb{R}^3$ latent space with new axes made from linear combinations of the training features, with the intent to separate the preassigned classes as much as possible. Our LDA model uses again the 129 features considered in the Debye temperature ML model, and is trained on 85\% of the samples from our superconductor dataset. After training, the model is then tested on the remaining 15\%.
Figure \ref{fig:LDA}(a) and \ref{fig:LDA}(b) left columns show the resulting data distributions on the first and second LD axes, respectively for the 85\% training and the 15\% test sets. The four manually selected classes of superconductors are found to be fairly well-separated in the latent space.
To test the classification quality, we resort to the confusion matrix, where the diagonal elements represent the numbers of correct class predictions, while the off-diagonal elements are those mislabeled by the classifier. 
As shown in Fig. \ref{fig:LDA}(a) and \ref{fig:LDA}(b) right columns, the confusion matrices (with the $y$-axis the actual class and the $x$-axis the predicted class) have large diagonal and small off-diagonal values, indicating a majority of correct predictions.

For a quantitative measure, we also use the class-weighted F1-score:
\begin{equation}
    F_1 = \displaystyle\sum\limits_{i=1}^{n} \frac{m_i}{M} \frac{\text{tp}}{\text{tp} + \frac{1}{2}(\text{fp} + \text{fn})}
\end{equation}
Here, $m_i$ is the number of samples in the $i$-th class, and $M (=\sum_{i=1}^n m_i$) is the total number of samples. tp (true positive), fp (false positive), and fn (false negative) are evaluated using the subset of samples in each class $i$.
The F$_1$-score ranges between 0 (meaning no correction classification) to 1 (meaning a perfect classifier). Our LDA ML model achieves high class-weighted F$_1$ scores of 0.973 for the training set [Fig. \ref{fig:LDA}(a)] and 0.980 for the test set [Fig. \ref{fig:LDA}(b)]. The results indicate that our classification scheme captures the essential differences between different families of superconductors based on the manual selection criteria and the features utilized to train the Debye temperature model. 

\begin{figure}[htp]
    \centering
    \includegraphics[width=0.48\textwidth]{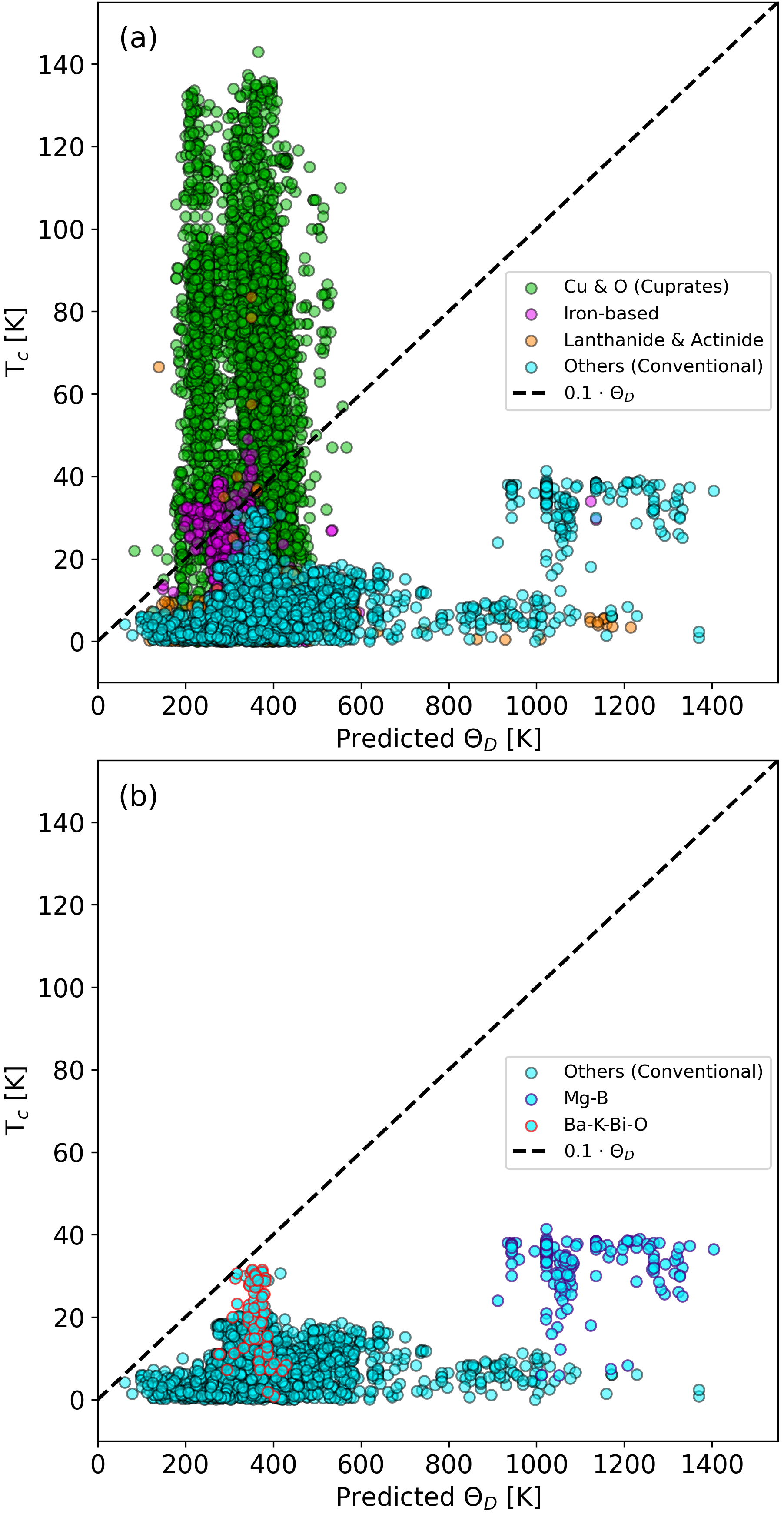}
    \caption{
    Scatter plots of experimental $T_c$ versus ML predicted Debye temperature  $\Theta_D$ for $\textbf{(a)}$ 9,860 selected superconductors (color-coded based on their class labels), and $\textbf{(b)}$ compounds classified as ``Others" in our manual selection criteria, with a majority of this class being BCS superconductors. The dashed line indicates the relationship $T_c = 0.1\Theta_D$. In $\textbf{(b)}$, the MgB$_2$ and the Ba$_{x}$K$_{1-x}$BiO$_{3}$ (BKBO) families of BCS superconductors are further highlighted by blue and red edge colors for discussion in the main text.
    }
    \label{fig:tc_debye}
\end{figure}

\subsection{Predicting Debye Temperature of Superconductors}
We next apply the ML models to predict the density and Debye temperature for our selected superconductors.
Figure \ref{fig:tc_debye}(a) shows the resulting scatter plot of the predicted $\Theta_D$ versus the experimental superconducting $T_c$ for all the 9,860 superconductors, which are further color-coded according to their class labels.
A linear dashed line with $T_c = 0.1\Theta_D$ is also plotted in Fig. \ref{fig:tc_debye} and serves as a guide to the eye.
It is seen that the compounds classified as the cuprates (Cu \& O) and iron-based superconductors have many entries that violate the bound $T_c\le 0.1 \Theta_D$.
While the majority of the lanthanide and actinide superconductors fall below this bound, they also have some exceptions with $T_c \gg 0.1 \Theta_D$. Overall, there is no clear trend between $T_c$ and $\Theta_D$ in the three superconductor families discussed above.
In Fig. \ref{fig:tc_debye}(b), we plot the relationship of $T_c$ versus $\Theta_D$ solely for compounds classified as ``Others" in our manual selection. This class contains 1,900 unique compounds with a majority being BCS superconductors.
The figure shows that \textit {all superconductors in this class fall in the proposed bound by Esterlis \textit{et al.}}~\cite{esterlis2018bound}

In the low Debye temperature regime ($\Theta_D \lesssim 400$ K) of Fig. \ref{fig:tc_debye}(b), some compounds reside in proximity to the plotted bound, but still their T$_{c}$ never exceeds $0.1 \Theta_{D}$.
Notably, the Ba$_{x}$K$_{1-x}$BiO$_{3}$ (BKBO) superconductors [highlighted by red edge color in Fig. \ref{fig:tc_debye}(b)] have a predicted $\Theta_D$ near 350 K, and their $T_c$ is very close to the proposed bound.
The relatively high BCS $T_c$ of BKBO has been attributed to enhanced electron-phonon couplings due to long-range Coulomb interaction (non-local screening) or electron correlation~\cite{Wen2018, li2019electron}. Therefore, exploring correlated BCS superconductors may help uncover new higher-$T_c$ compounds. On the other hand, in the higher Debye temperature regime ($\Theta_D \gtrsim 400$ K), the experimental $T_c$ is far away from the bound. For example, MgB$_2$ [highlighted by blue edge color in Fig. \ref{fig:tc_debye}(b)] is a BCS superconductor with the highest known T$_c\sim$ 39K-42K \cite{Nagamatsu2001, Pallecchi2005} at ambient pressure. Its $\Theta_D\sim 1000K$ results in a $T_c/\Theta_D$ ratio of only $\sim 0.04$. Therefore, there is a possibility of engineering electron-phonon or other properties to further enhance $T_c$ for relatively high-$\Theta_D$ BCS superconductors, via for example strain, doping, and/or external pressure.

\begin{figure*}[ht]
    \centering
    \includegraphics[width=0.97\textwidth]{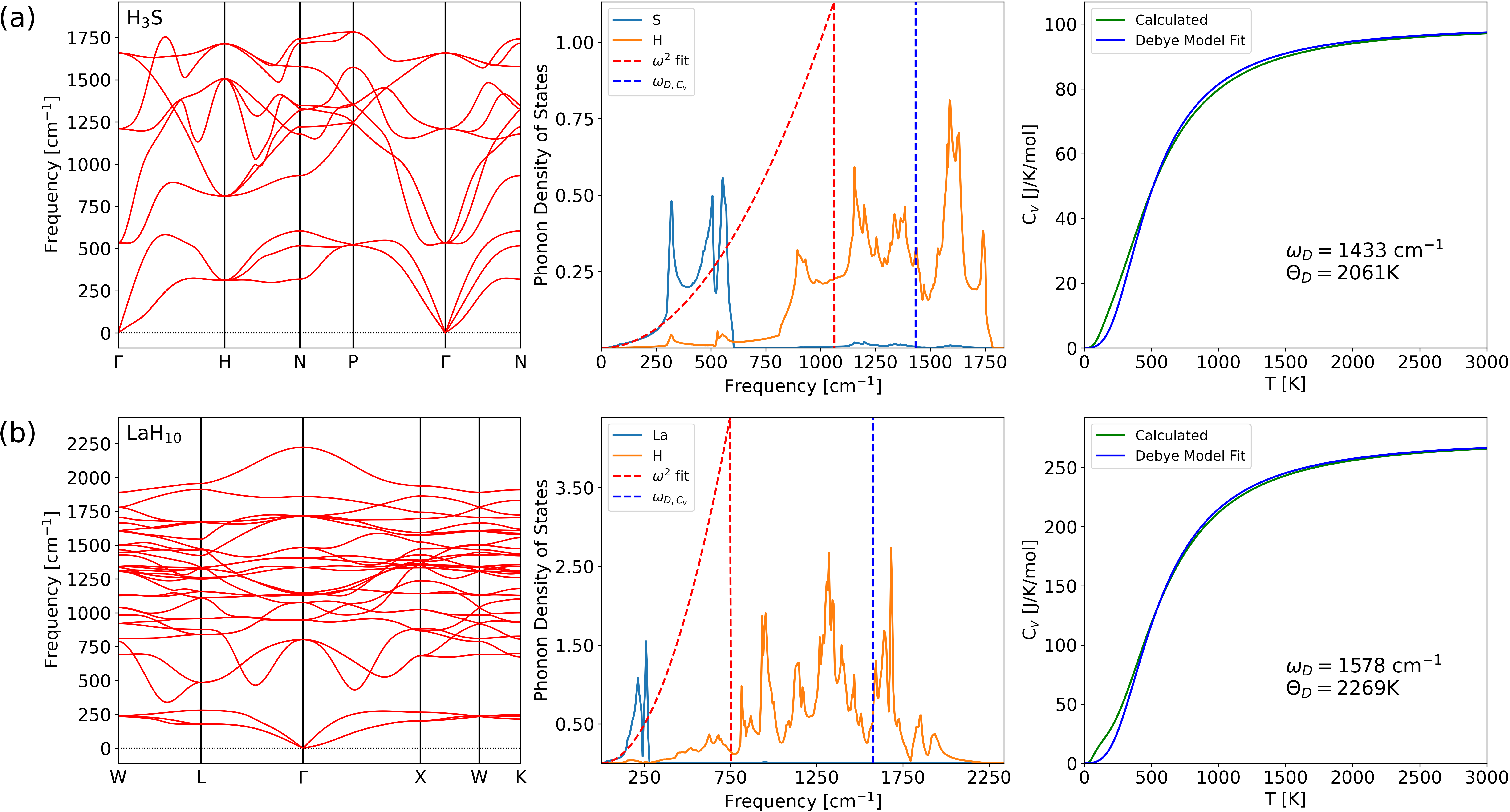}
    \caption{
    First-principles phonon calculations and Debye temperature analysis for \textbf{(a)} H$_{3}$S ($Im$-$3m$) and \textbf{{(b)}} LaH$_{10}$ ($Fm$-$3m$), under 200 GPa.
    The left panel is the phonon dispersion; the middle panel is the phonon density of states (DOS) projected onto each atom; the right panel is the specific heat from the phonon calculations and its fit to the Debye model.
    As discussed in the main text, the Debye frequency $\omega_D$ determined by a quadratic fit to the low-energy phonon modes and the phonon DOS is shown as the red dashed line in the middle panel; $\omega_D$ determined by a fit to the Debye model is indicated by the blue dashed line. The Debye temperature is then calculated as $\Theta_{D}=\hbar\omega_{D}/k_B$. 
    }
    \label{fig:HP_hydrides}
\end{figure*}

\subsection{First-Principles Phonon Calculations of High-Pressure Hydrides}

Finally, we discuss the relationship between $T_c$ and $\Theta_D$ in the recently discovered high-pressure hydride (BCS) superconductors. In particular, we focus on H$_{3}$S and LaH$_{10}$ under an external pressure of 200 GPa, which have experimentally reported $T_c \sim 203$ K~\cite{Drozdov2015} and $T_c\sim 250-260$ K~\cite{Drozdov2019, Somayazulu2019}, respectively.
Since our machine learning models are not applicable for high pressure studies, here we resort to first-principles calculations.
We note that Eq. \ref{eq:1} based on the mean sound velocity and mechanical properties can substantially underestimate $\Theta_D$ for systems near a structural phase transition. For example, the superconducting sulfur hydride undergoes a structural transition from $R3m$ to $Im$-$3m$ symmetry near 150 GPa~\cite{einaga2016crystal}. 
In lanthanum hydride, the most stable structure near 200 GPa is of $Fm$-$3m$ symmetry, but there also exist several different phases of similar enthalpies~\cite{errea2020quantum}.
In these cases, the elastic constants computed in the pressure range $\sim 150-200$ GPa may exhibit a significant softening behavior. Therefore, instead of using Eq. \ref{eq:1}, we perform direct first-principles phonon calculations to estimate the Debye temperature from the phonon density of states (DOS) and the associated specific heat~\cite{parlinski1997first, Chaput2011, Togo2015}.

Figures \ref{fig:HP_hydrides}(a) and \ref{fig:HP_hydrides}(b) left columns show respectively the phonon dispersion spectra for H$_{3}$S ($Im$-$3m$) and LaH$_{10}$ ($Fm$-$3m$) under 200 GPa. The spectra exhibit no negative phonon modes, indicating dynamical stability of the underlying crystal structure.
The phonon bandwidths are $\sim 1750$ cm$^{-1}$ for H$_{3}$S and $\sim 2250$ cm$^{-1}$ for LaH$_{10}$, which may serve as a (very rough) estimation of $\Theta_D$. A more accurate $\Theta_D$ can be determined from the phonon DOS shown in Fig. \ref{fig:HP_hydrides} middle panels.
In both H$_3$S and LaH$_{10}$, the hydrogen projected DOS (PDOS) lie at higher energy and are well separated from the lower-energy PDOS of sulfur or lanthanum. Because of this clear separation, a typical $\omega^{2}$ fit (with $\omega$ the phonon frequency) using just the low-frequency phonon modes and the area underneath the phonon DOS would underestimate $\Theta_{D}$. As indicated by the red dashed lines in Fig. \ref{fig:HP_hydrides} middle panels, the $\omega^{2}$ fit leads to significantly smaller values of $\Theta_{D}=1575$ K (or Debye frequency $\omega_D = 1054$ cm$^{-1}$) for H$_3$S and $\Theta_{D}=1070$ K ($\omega_D = 743$ cm$^{-1}$) for LaH$_{10}$.
These values are clearly smaller than the expected first-moment average of the phonon energies $\langle \omega \rangle$, which in our calculations are 1187 cm$^{-1}$ and 1294 cm${-1}$ respectively for H$_3$S and LaH$_{10}$. Therefore, a more sensible and accurate alternative method to evaluate $\Theta_D$ based on the phonon spectra should be considered.

In systems where the phonon DOS show an apparent frequency gap or exhibit multiple branches, one can determine $\Theta_D$ by fitting the computed $c_v$ (specific heat at constant volume) from the phonon DOS directly to that in the Debye model: 
\begin{equation}\label{Debye_model_cv}
    c_v =9 N k_{B}\left(\frac{T}{\Theta_{\mathrm{D}}}\right)^{3} \int_{0}^{\Theta_{\mathrm{D}} / T} \frac{x^{4} e^{x}}{\left(e^{x}-1\right)^{2}} d x.
\end{equation}
Here, $N$ is the number of atoms in the unit cell multiplied by the Avogadro's number, and $k_{B}$ is the Boltzmann constant.
Figure \ref{fig:HP_hydrides} right panels show the computed $c_v$ from the phonon DOS and its fit to the Debye model.
The fitting procedures lead to $\Theta_{D}=2,061$ K for H$_3$S and $\Theta_{D}=2,269$ K for LaH$_{10}$, which are more consistent and expected from a weighted average of the phonon energies.
The results indicate that $T_c/\Theta_D \sim 0.10$ for H$_{3}$S and 0.11 for LaH$_{10}$. Therefore, these high-pressure hydrides essentially saturate the proposed bound by Esterlis \textit{et al.}~\cite{esterlis2018bound}
It would be an interesting and important future study to verify if other high-pressure hydrides such as CaH$_6$~\cite{ma2022high} and LaBeH$_8$~\cite{song2023stoichiometric} also have a $T_c/\Theta_D$ ratio that either falls within or resides close to the proposed bound.

\section{Conclusion}
We have developed machine learning models to predict the Debye temperature $\Theta_D$ for 9,860 superconductors and compared the values to their experimental superconducting transition temperature $T_c$.
In our dataset, all the conventional phonon-mediated superconductors (nearly 2,000 compounds) were found to satisfy the relationship $T_c\le A\Theta_D$, with $A\sim 0.1$~\cite{esterlis2018bound}. We also have performed first-principles phonon calculations to investigate H$_3$S and LaH$_{10}$ under 200 GPa, and found that $T_c \sim 0.1$ to 0.11 $\Theta_D$ in these high-pressure hydrides.
Therefore, these results imply that while $T_c$ might be further engineered by external perturbations, it is {\it unlikely} that $T_c/\Theta_D$ can be made much larger than 0.1 in BCS superconductors.

Currently, our machine learning model takes only compositional features and crystal system symmetry as input. The model has not been trained using other structural information such as lattice parameters or atomic positions, which will change with pressure. A more sophisticated model architecture like the Crystal Graph Convolutional Neural Network (CGCNN)~\cite{xie2018crystal} can take an arbitrary crystal structure and learn from the connection of atoms to predict material properties. Therefore, it could be an interesting future direction to train CGCNN-like models that take pressure and/or structural information as input for predicting BCS superconductors under pressure. One limitation, however, is that high-pressure data remain scarce in existing materials databases, so it is also important to construct sufficient training data of compressed crystal structures across a desired pressure range in the future.

Our study also suggests a few possibilities to enhance $T_c$ of phonon-mediated superconductors. For example, one may push $T_c$ towards the bound by an enhanced electron density of states or electron-phonon couplings, due to interaction or non-local screening effects in correlated (BCS) superconductors, such as Ba$_{x}$K$_{1-x}$BiO$_{3}$. For existing materials with a high $\Theta_D$, their $T_c$ is in general far away from the bound, so there should be opportunity for enhancing $T_c$ by engineering the electron-phonon or other properties, via doping, alloying, strain, or pressure. Finally, achieving room-temperature BCS superconductors requires a corresponding high Debye temperature, which can occur under high pressure, or in materials with a high lattice thermal conductivity (high phonon frequency) or superior mechanical properties (large elastic constants). Machine learning discovery of these materials such as superhard metals~\cite{chen2021machine} for new BCS superconductors might be interesting areas of future studies.

\section*{Acknowledgments}
The authors thank Steven Kivelson for fruitful discussion. A.D.S., R.P.C., and C.-C.C. are support by the National Science Foundation (NSF) RII Track-1 Future Technologies and Enabling Plasma Processes (FTPP) Project OIA-2148653.
A.D.S acknowledges support from the UAB Blazer Fellowship and the FTPP CERIF Graduate Research Assistantship.
C.-C.C. is also supported by NSF award DMR-2142801.
S.B.H. is supported by the Center for Nanophase Materials Sciences (CNMS), which is a US Department of Energy, Office of Science User Facility at Oak Ridge National Laboratory.
D.Y. acknowledges support from ARDEF 1ARDEF21 03 from ADECA and NSF award OAC-2106461.
Part of the calculations were performed on the Frontera computing system at the Texas Advanced Computing Center. Frontera is made possible by NSF award OAC-1818253. 



\bibliography{main_v3}

\end{document}